\documentstyle[12pt]{article}
\def\one{1\hskip-.37em 1}     
\def\l{\lambda}
\def\D{{\cal D}}
\def\K{{\cal K}}
\def\E{{\rm E}\hskip-.55em{\rm I}}
\def\ir{{\rm I}\hskip-.2em{\rm R}}
\def\half{\textstyle{\frac{1}{2}}}
\def\quarter{\textstyle{\frac{1}{4}}}

\def\ra{\rightarrow}
\def\tint{{\textstyle{\int}}}
\def\d{\partial}
\def\o{\overline}
\def\b{\begin{eqnarray*}}     
\def\e{\end{eqnarray*}}       
\def\bn{\begin{eqnarray}}     
\def\en{\end{eqnarray}}       
\def\<{\langle}
\def\>{\rangle}

\def\{{\lbrace}
\def\}{\rbrace}

\title{Coherent State Quantization \\of Constraint Systems}
\author{John R. Klauder\\
Departments of Physics and Mathematics\\
University of Florida\\
Gainesville, Fl  32611\\
and\\
I.H.E.S.\\
F-91440 Bures-sur-Yvette}
\begin{document}
\maketitle
\begin{abstract}
 A careful reexamination of the quantization of systems with first- and second-class constraints from the 
point of view of coherent-state 
phase-space path integration reveals several significant distinctions from more conventional treatments. 
Most significantly, we emphasize the importance of using path-integral measures for Lagrange multipliers 
which ensure that the quantum system satisfies the quantum constraint conditions. Our procedures involve 
no $\delta$-functionals of the classical constraints, no need for gauge fixing of first-class constraints, no 
need to eliminate second-class constraints, no potentially ambiguous determinants, and have the virtue of 
resolving differences between canonical and path-integral approaches. Several examples are considered in 
detail.
\end{abstract}
\section{Introduction}
The traditional quantization of systems with first-class constraints by means of conventional phase-space 
path integrals entails $\delta$-functionals to enforce the constraints, additional $\delta$-functionals to 
choose a gauge, as well as a determinant brought about as the Jacobian of a change from canonical to 
noncanonical variables \cite{fad}. In various applications, such as non-Abelian gauge theories, the chosen 
gauge fixing may be nonadmissable and one form or another of ambiquity may arise in such a 
quantization procedure \cite{gri,gov}. Most commonly, second-class constraints are solved for and 
eliminated from the theory before quantization, but there is another procedure in which $\delta$-
functionals and a determinant of their own appear in the path integral expression \cite{sen}. In this paper 
we reexamine phase-space path integral quantization from a coherent-state viewpoint, and arrive at 
procedures that (i) do not involve gauge fixing and hence require no potentially ambiguous determinant in 
the case of first-class constraints, and (ii) do not require either the elimination of any variables or the 
introduction of Dirac brackets \cite{dir} in the case of second-class constraints.

We first observe that the classical action functional (summation implied)
  \bn I=\int[p_j\,{\dot q}^j-H(p,q)-\l^a\phi_a(p,q)]\,dt\;,  \en
$1\leq j\leq J$, and $1\leq a\leq A\leq 2J$,
leads both to Hamilton's equations
 \bn    {\dot q}^j&=&\frac{\d H}{\d p_j}+\l^a\,\frac{\d \phi_a}{\d p_j}\;,\\
    {\dot p}_j&=&-\frac{\d H}{\d q^j}-\l^a\,\frac{\d\phi_a}{\d q^j}   \en
as well as the constraint equations 
   \bn \phi_a(p,q)=0,\hskip2cm 1\leq a\leq A\;,\en
 obtained by varying the Lagrange multipliers $\{\l^a\}$. The entire theory, dynamics included, is thus 
forced to lie on the {\it constraint hypersurface} defined as that submanifold in phase space specified by 
the conditions $\phi_a(p,q)=0$, $1\leq a\leq A$. 
In order to remain on the constraint hypersurface it is necessary that
 \bn {\dot\phi}_a(p,q)=\{\phi_a(p,q),H(p,q)\}+\l^b\{\phi_a(p,q),\phi_b(p,q)\}\equiv0\;. \en
At this point we assume that the original set of constraints is complete in the sense that no new constraints 
are needed to satisfy (5). 
Given that assumption, it is useful to divide constraints into two classes \cite{dir}.
First-class constraints are distinguished by the fact that their Poisson brackets satisfy 
  \bn  \{\phi_a(p,q),\,\phi_b(p,q)\}&=&c_{ab}^{\;\;\;\;c}\,\phi_c(p,q)\;,  \\
   \{\phi_a(p,q),H(p,q)\}&=&h_a^{\;\;b}\,\phi_b(p,q)\;,   \en
and hence these Poisson brackets vanish on the constraint hypersurface. This conclusion holds even if the 
quantities $c_{ab}^{\;\;\;\;c}$ and $h_a^{\;\;b}$ are suitable functions of the phase-space variables 
themselves, however for the present we will assume that the Poisson bracket algebra of the constraints 
leads to a closed algebra, namely that the quantities $c_{ab}^{\;\;\;\;c}$ and $h_a^{\;\;b}$ are simply 
constants leading, in particular, to the conclusion that the constraints form a Lie algebra with the Lie 
bracket given by the Poisson bracket. For first-class constraints it follows---from the bracket structure of 
the constraints with themselves and with the Hamiltonian---that the first time derivative of the constraint 
vanishes on the constraint manifold without any further conditions; this also holds for all higher-order 
time derivatives as well. Thus it suffices to impose the constraints at the initial time, say $t=0$, and 
thereafter the canonical equations of motion will ensure that the constraints are satisfied for all $t>0$. 
Such an imposition of the constraints at $t=0$ is called an {\it initial value equation}. It is typical of 
first-class constraints that the associated Lagrange multipliers are {\it not} determined by the solution to 
the equations of motion, but rather the temporal behavior of the Lagrange multipliers must be specified in 
advance so as to determine a unique solution to the canonical equations of motion. This arbitrary but 
classically necessary choice of the Lagrange multipliers is referred to as a {\it choice of gauge}. No 
variable that is deemed to be observable should depend on the arbitrary choice of the Lagrange multipliers 
in this case, and that rule excludes gauge-variable quantities from being observables. For systems with 
first-class constraints, only quantities $F$ for which 
\bn \{F,\phi_a(p,q)\}=f_a^{\;\;b}\phi_b(p,q)  \en 
for some (possibly phase-space dependent) variables $f_a^{\;\;b}$ qualify as observables. For such 
quantities we observe that
  \bn{\dot F}=\{F,H\}+\l^a\{F,\phi_a\}=\{F,H\}+\l^af_a^{\;\;b}\phi_b \en
has a vanishing contribution from the Lagrange multipliers on the constraint hypersurface as needed for 
their independence of the arbitrary time dependence of $\{\l^a\}$.

If Eq.~(6) fails to hold, then we say we are dealing with second-class constraints. In that case some or all 
of the Lagrange multipliers are explicitly determined in order that (5) be satisfied.  We will take up the 
question of second-class constraints at a later stage, and for the present we assume that our constraints are 
all first class.
\subsubsection*{Conventional Hamiltonian quantization}
The standard way of approaching systems with first-class constraints, following Dirac \cite{dir}, 
quantizes first and imposes the constraints as a second step. In particular, we assume that the original 
classical phase-space variables are already expressed in suitable phase-space coordinates so that we may 
``promote'' them to canonical, irreducible, self-adjoint operators $Q^j$ and $P_j$, $1\leq j\leq J$, such 
that the only nonvanishing commutation relation is
  \bn [Q^j,P_k]=i\delta^j_k\one  \en
in units for which $\hbar=1$. A self-adjoint operator ${\cal H}(P,Q)$ serves as the Hamiltonian, and as 
usual leads to the Heisenberg equations of motion
  \bn &&{\dot Q}^j=i[Q^j,{\cal H}]\;,\\
    &&{\dot P}_j=i[P_j,{\cal H}]\;.  \en
To impose the constraints, we enforce the initial value equation at $t=0$ in the form
  \bn  \Phi_a(P,Q)\,|\psi\>_{\rm phys}=0\;,  \en
where $|\psi\>_{\rm phys}$ represents a vector in the physical Hilbert space. Consistency of this equation 
is ensured by the requirement that
  \bn  &&[\Phi_a(P,Q),\Phi_b(P,Q)]=ic_{ab}^{\;\;\;\;c}\,\Phi_c(P,Q)\\
  &&[\Phi_a(P,Q),{\cal H}(P,Q)]=ih_a^{\;\;b}\,\Phi_b(P,Q)\;.  \en
These relations guarantee that the constraint conditions hold for all time once they hold at the initial time. 

The Dirac procedure sketched above is unsatisfactory whenever some of the constraint operators $\Phi_a$ 
possess a {\it continuous spectrum}; in that case it is impossible to find normalized vectors, i.e., Hilbert 
space vectors, as solutions to the proposed initial value condition. In the procedure that we advance in this 
paper we shall also have to face the problem of constraint operators with continuous spectrum, and we 
must come up with some way of dealing with them \cite{lan}. 

There is also another important way of quantizing systems with first-class constraints, namely by path 
integration, which we examine next. 
\subsubsection*{Conventional phase-space path integration}
The conventional phase-space path integration for systems with closed first-class constraints \cite{fad} 
assumes at the outset that in performing the integration over the Lagrange multipliers, the measure on 
$\{\l^a\}$ is flat and that the range of integration at each time instant is $\ir^A$, all this in order to secure 
$\delta$-functionals of the constraints. In customary formal notation for phase-space path integrals, the 
first step in quantization is therefore given by
  \bn &&\int\exp\{i\tint[p_j{\dot q}^j-H(p,q)-\l^a\phi_a(p,q)]\,dt\}\,\D p\,\D q\,\D\l\nonumber\\
 &&\hskip1cm =\int\exp\{i\tint[p_j{\dot q}^j-H(p,q)]\,dt\}\delta\{\phi(p,q)\}\,\D p\,\D q\;.   \en
This result of the Lagrange multiplier integration ensures that the subsequent domain of integration is 
confined to the classical constraint hypersurface, but it also generally leads to divergences in the 
remaining integrations due to the independence of the integrand on those variables conjugate to the 
constraint variables. To tame such divergences, $A$ auxiliary conditions, $\chi^a(p,q)=0$, generally 
referred to as a gauge choice, are adopted through the introduction of additional $\delta$-functionals, 
leading to the formal expression
 \bn \int\exp\{i\tint[p_j
{\dot q}^j-H(p,q)]\,dt\}\delta\{\chi(p,q)\}\,\det\{\chi^a,\phi_b\}\,\delta\{\phi(p,q)\}\,\D p\,\D q\;, \en
where the determinant arises as the Jacobian of the transformation from canonically conjugate conjugates 
to the constraints to the simply conjugate choices represented by the $\{\chi^a\}$. For various sets 
$\{\xi^a\}$, the expressions $\xi^a\phi_a(p,q)$ serve as generators of gauge transformations, and lead to 
gauge-equivalent sets (fibers) in the constraint hypersurface. It is assumed that the gauge conditions 
$\{\chi^a=0\}$ pick out a single unique representative from each and every equivalence class (a section 
from each fiber). Ideally, the determinant is nowhere vanishing, but that is not the only condition that 
needs to be fulfilled; a thorough discussion of these points is available in the monograph of Govaerts 
\cite{gov}. Under certain circumstances, rather natural appearing gauge condition choices are actually 
nonadmissable and the procedure outlined above may not hold. Although this situation can be a serious 
problem, such issues are not our primary concern, and we shall not dwell on such potential ambiquities 
but instead refer to the literature \cite{gri,gov}. 

Let us assume for the moment that the formal procedures outlined above are acceptable from the point of 
view of the choice of gauge. Then one may proceed to the next step in which the two sets of $\delta$-
functionals are used to eliminate $2A$ variables leading to a formal path integral expression on the {\it 
reduced phase space} consisting of $2(J-A)$ variables $\{p^*_r,q^{*\,r}\}$ for which the last expression 
(17) for the propagator is transformed into
\bn\int\exp\{i\tint[p^*_r{\dot q}^{*\,r}-H^*(p^*,q^*)]\,dt\}\,\D p^*\,\D q^*\;;  \en
here $H^*(p^*,q^*)$ represents the reduced Hamiltonian in which the constraints and gauge conditions 
have been imposed to lead to this final expression. 

Our concern with the final formal path integral expression on the reduced phase space holds even in the 
absence of any special problems that may arise when fixing the gauge. In particular, we may wish to go 
beyond the formal path integral that this expression represents and attempt to calculate some quantity, 
even approximately. For that purpose a suitable definition of the formal path integral needs to be given, 
and despite formal canonical coordinate covariance of the expressions involved, path integrals are well 
known to be ambiguous and essentially undefined {\it ab initio} when formulated in curvilinear phase-
space coordinates as is generally the case on the reduced phase space. Moreover, the proper lattice 
definition for conventional phase-space path integrals always has one more momentum integration than 
coordinate integration. This discrepancy is not serious for questions of dynamics, for which the deviation 
from unity is small for each time slice, but it is significant for imposing constraints inasmuch as the 
deviation from unity for each time slice for constraints is not small, needing, for example, to impose a 
$\delta$-function of the classical constraint conditions. In addition, an unequal number of momentum and 
coordinate integrations is highly incompatible with general canonical transformations needed and often 
used in formal discussions of conventional phase-space path integrals.

\subsubsection*{Outline of the paper}
We wish to reexamine the question of quantization of constraint systems, both first class as well as second 
class, and in that process we shall argue that the imposition of $\delta$-functionals may well not be the 
appropriate way to enforce that the system is confined to  a {\it quantum constraint subspace}. We shall 
employ a different integration measure for the Lagrange multipliers that is specifically designed to enforce 
the quantum constraint conditions rather than the classical constraint conditions. Coherent-state 
techniques, which are briefly reviewed, are found to be very natural in our development of an alternative 
path-integral quantization for systems with constraints. Most importantly, we shall find that even though 
the quantum constraints force the system to ``live'' in a reduced Hilbert space, the inner product is 
generally given with a measure having the original number of integration variables as well as the original 
domain of integration. Thus there are no $\delta$-functionals appearing in our formulation. In addition, 
for systems with closed first-class constraints, as we shall see, the quantization is {\it already} manifestly 
gauge invariant. Several examples of coherent-state path integral quantization of systems with first-class 
and second-class constraints are given, as well as a brief discussion of a recent toy model \cite{lee} of 
non-Abelian gauge fields.
\section{Coherent States}
Path-integral quantization has always been best understood as a construct from a well-defined operator 
formulation, and this dictum is no less true in the case of coherent-state path integrals. Canonical coherent 
states are  defined (in our phase convention) by the expression
  \bn |p,q\>\equiv\exp{[i\alpha(p,q)]} 
\exp(-iq^j\,P_j)\,\exp(ip_j\,Q^j)\,|\eta\>\;,\hskip1cm\alpha(p,q)\equiv0\;, \en
and involve irreducible, self-adjoint canonical Heisenberg operators that satisfy the standard commutation 
relation (10), as well as some normalized fiducial vector $|\eta\>$. For any choice of $|\eta\>$, it follows 
that these states admit a resolution of unity given as a superposition of one-dimensional projection 
operators by
\bn \one=\int|p,q\>\<p,q|\,d\mu(p,q)\;,\;\;\;\;\;\;\;\;\;\;d\mu(p,q)\equiv d^J\!p\,d^J\!q/(2\pi)^J\;, \en
where the measure $\mu$ is uniquely given as shown and the domain of integration is the entire space 
$\ir^{2J}$. An important choice of the fiducial vector is the ground state of an harmonic oscillator, say 
one with unit angular frequency, and it will be convenient to use such a fiducial vector for illustrative 
purposes. For this fiducial vector it follows, for a single degree of freedom ($J=1$), for example, that  
\bn\hskip-.5cm\<p'',q''|p',q'\>\!\!\!\!&=&\!\!\!\!\pi^{-1/2}\int_{-\infty}^\infty \exp[-\half(k-p'')^2+ik(q''-q')-
\half(k-p')^2]\,dk\nonumber\\
&=&\!\!\exp\{i\half(p''+p')(q''-q')-\quarter[(p''-p')^2+(q''-q')^2]\}\,.\en
For multiple degrees of freedom the exponent becomes a sum of similar expressions.

Expectation values in the coherent states are of considerable interest. In particular, if ${\cal H}(P,Q)$ 
denotes a self-adjoint operator, say the quantum Hamiltonian for concreteness, then the diagonal 
expectation
  \bn \<p,q|{\cal H}(P,Q)|p,q\>\equiv H(p,q)=\<p,q|:H(P,Q):|p,q\>\en
defines a real function of $p$ and $q$ that we will adopt as the classical Hamiltonian in a formulation in 
which the classical and quantum theories {\it coexist} (as they do in the real world!). Observe that the 
relation of $\cal H$ with $H$ is that of a normal-ordered expression (with respect to the fiducial vector) as 
given in the last part of (22). As special cases note that
  \bn &&\<p,q|P_j|p,q\>=p_j\;,  \\
     &&\<p,q|Q^j|p,q\>=q^j\;,  \en
and since $p_j$ and $q^j$ are expectation values (as opposed to eigenvalues) there is no conflict in 
specifying both of them simultaneously. 

An important one form determined by the coherent states is given by
\bn  i\<p,q|d|p,q\>=p_j\,dq^j\;,   \en
and this form holds for the wide class of fiducial vectors that lead to (23) and (24).
This expression enters into the construction of the coherent-state path integral as we shall now show.
\subsubsection*{Coherent-state path integral}
The coherent-state matrix elements of the evolution operator are defined as the propagator, namely, 
  \bn   \<p'',q''|\,e^{-i{\cal H}T}\,|p',q'\>\;,  \en
which may be rewritten [with $\epsilon\equiv T/(N+1)$] as 
  \bn  &&\hskip-1cm\<p'',q''|e^{-i{\cal H}\epsilon}e^{-i{\cal H}\epsilon}\cdots 
e^{-i{\cal H}\epsilon}|p',q'\>\nonumber\\
&&\hskip-1cm =\int\prod_{l=0}^N\<p_{l+1},q_{l+1}|
e^{-i{\cal H}\epsilon}|p_l,q_l\>\,\prod_{l=1}^N\,dp^J_l\,dq^J_l/(2\pi)^J\nonumber\\ &&\hskip-1cm 
=\lim\int\prod_{l=0}^N\{\<p_{l+1},q_{l+1}|p_l,q_l\>-i\epsilon\<p_{l+1},q_{l+1}|
{\cal  H}|p_l,q_l\>\}\,\prod_{l=1}^N\,d\mu(p_l,q_l)\;;
\en
here we have introduced the limit $\epsilon\ra0$, as well as the notation $p_{N+1},q_{N+1}\equiv p'',q''$ 
and $p_0,q_0\equiv p',q'$, {\it which we shall also consistently use hereafter without further comment}.

Additionally, we recall the formal path integral expression that arises when the order of the limit and the 
integrations are interchanged, and the integrand is written in the form it assumes for continuous and 
differentiable paths. In such a formal procedure, the propagator reads
  \bn &&\hskip-1cm\int\exp\{i\tint[i\<p,q|\frac{d}{dt}|p,q\>-\<p,q|
{\cal H}|p,q\>]\,dt\}\,\D\mu(p,q)\nonumber\\
&&={\cal M}\int\exp\{i\tint[p_j{\dot q}^j-H(p,q)]\,dt\}\,\D p\,\D q\;,  \en
where $\cal M$ denotes a formal normalization factor (coming from the ``$1/2\pi$'' terms).

We emphasize that the final expression (28)---no matter how appealing it may seem---is heuristic and 
formal. It provides only the {\it classical entry level} to the calculation and does not by itself suggest a 
definition that is guaranteed to be correct. One has only to pass from one set of canonical coordinates to 
another to convince oneself that the formal path integral in Eq.~(28) is inadequately defined and not ready 
to evaluate as it stands. The value of such an equation lies not in its calculational readiness, but rather in 
its accessibility to classical perceptions. 
\section{Reproducing Kernel Hilbert Spaces}
It is pedagogically useful at this point for us to review the kind of Hilbert space representation generated 
by coherent states.\footnote{Readers more interested in applications to constraint systems may wish to 
proceed directly to Sec.~5 on a first reading.} Consider arbitrary matrix elements of the resolution of 
unity, Eq.~(20), which read
  \bn \<\phi|\psi\>=\tint\<\phi|p,q\>\<p,q|\psi\>\,d\mu(p,q)\;.  \en
It follows, therefore, that $\psi(p,q)\equiv\<p,q|\psi\>$ defines a phase-space representation of Hilbert 
space, and moreover it is composed of square-integrable functions all of which are {\it bounded and 
continuous}. The inner product in this functional space is therefore given by
  \bn  (\phi,\psi)\equiv \tint \phi^*(p,q)\psi(p,q)\,d\mu(p,q)\;, \en
for all functions in the space. The resultant Hilbert space does {\it not} consist of {\it all} square 
integrable functions of the indicated variables, but only a closed subspace of such functions. We may 
characterize this subspace as follows. Let us set $\<\phi|=\<p'',q''|$, namely one of the coherent states, 
which leads to 
  \bn \<p'',q''|\psi\>=\tint\<p'',q''|p,q\>\<p,q|\psi\>\,d\mu(p,q)\;.\en
This equation asserts that each element of the subspace is {\it reproduced} by means of a {\it reproducing 
kernel} \bn  \K(p'',q'';p,q)\equiv\<p'',q''|p,q\>\;,  \en
 which is just the coherent-state overlap function. If $\K$ were a $\delta$-function there would be no 
information in an equation such as (31); since $\K$ is a bounded and continuous function, there is, in fact, 
a great deal of information in Eq.~(31). The reproducing kernel satisfies an integral equation of its own 
obtained if we further specialize $|\psi\>=|p',q'\>$, from which we learn that
 \bn  \<p'',q''|p',q'\>=\tint\<p'',q''|p,q\>\<p,q|p',q'\>\,d\mu(p,q)\;. \en
This integral equation, along with the fact that $\<p,q|p',q'\>^*=\<p',q'|p,q\>$, establishes that the 
reproducing kernel serves as a projection operator in the space of all square-integrable functions onto the 
subspace of interest. 

It is important to appreciate that {\it the entire functional Hilbert space is fully characterized by the 
reproducing kernel} $\K$. In particular, a dense set of elements in the Hilbert space is given by functions 
of the form
  \bn  \psi(p,q)\equiv\sum_{k=1}^K\alpha_k\<p,q|p_k,q_k\>  \en
for complex coefficients $\{\alpha_k\}$, phase-space points $\{(p_k,q_k)\}$, and a finite sum, $K<\infty$. 
If another such vector is given by
  \bn  \phi(p,q)\equiv\sum_{l=1}^ L\beta_l\<p,q|{\o p}_l,{\o q}_l\>\;,\en
based on the phase-space points $\{({\o p}_l,{\o q}_l)\}$, then it follows for any pair of such vectors (both 
of which lie in a dense set) that 
  \bn (\phi,\psi)=\sum_{l=1}^{L}\sum_{k=1}^{K}\beta^*_l\alpha_k\<{\o p}_l,{\o q}_l|p_k,q_k\>\;.\en
Completion of this space is given by including limit points, all of which are bounded continuous 
functions, and are limits of Cauchy sequences in the norm $\|\psi\|$ defined for vectors in the dense set by
  \bn \|\psi\|^2\equiv\sum_{k=1}^{K}\sum_{l=1}^{K}\alpha^*_k\alpha_l\<p_k,q_k|p_l,q_l\>
\;.\en 
The reproducing kernel is unique and therefore this function entirely characterizes the functional Hilbert 
space all by itself.  If a given nonzero function belongs in two reproducing kernel Hilbert spaces, then the 
two spaces are not merely equivalent, they are {\it identical}. Change the reproducing kernel and---apart 
from the zero element $\psi(p,q)\equiv0$---{\it every functional representative is changed}. 

Indeed, it is instructive to set aside the bra-ket notation temporarily and note that {\it any} bounded, 
continuous function of two ``phase-space'' sets, i.e.,
$\K(p'',q'';p',q')$, which satisfies the two basic properties that
  \bn &&\K(p',q';p'',q'')=\K(p'',q'';p',q')^*\;,\\ &&\sum_{k=1}^{K}\sum_{l=1}^{K}\alpha^*_k\alpha_l\K(p_k,q_k;p_l,q_l)\,\geq0\en
for all sets $\{\alpha_l\}$, $\{(p_l,q_l)\}$, and all finite $K$, is an acceptable reproducing kernel defining 
a reproducing kernel Hilbert space \cite{mes}. A dense set of vectors in the associated Hilbert space is 
given by functions of the form
  \bn  \psi(p,q)=\sum_{l=1}^K\alpha_l\,\K(p,q;p_l,q_l)\;,  \en
and the inner product of two such vectors is {\it defined} by
  \bn  (\phi,\psi)=\sum_{l=1}^{L}\sum_{k=1}^{K}\beta^*_l\alpha_k\K({\o p}_l,{\o q}_l;p_k,q_k)\;. \en
What is different in the present case, is that there is no guarantee---{\it nor, in the general case, any 
requirement}---that the inner product also admits a local integral representation in the form
 \bn  (\phi,\psi)=\tint\phi^*(p,q)\psi(p,q)\,d\rho(p,q)  \en
for some positive measure $\rho$. On the other hand, coherent states and their associated resolution of 
unity lead to reproducing kernel Hilbert spaces that do have an alternative inner product definition in the 
form of a local integral with a positive measure.

It is important to add, according to the G.N.S. (Gel'fand, Naimark, Segal) Theorem \cite{gns}, that every 
bounded and continuous kernel function that satisfies Eqs.~(38) and (39) admits a representation as the 
inner product of two vectors in a (separable) Hilbert space. In other words, there is a Hilbert space and a 
set of vectors $\{\Phi[p,q]\}$ therein for which
  \bn  (\Phi[p'',q''],\Phi[p',q'])\equiv\K(p'',q'';p',q')\;.  \en
These vectors are uniquely determined by the kernel itself apart from unitary equivalence, i.e., whether the 
vectors $\Phi[p,q]$ are represented in $l^2$ or in $L^2$, etc. Since all such representations are unitarily 
equivalent we can, without loss of generality, simply choose the reproducing kernel Hilbert space itself as 
our representation space. 
In this generality note that there is no requirement of normalization of the vectors $\Phi[p,q]$. We further 
observe that if one reproducing kernel is proportional to another reproducing kernel, e.g., 
$\K_c(p'',q'';p',q')=c\K(p'',q'';p',q')$, $0<c<\infty$, $c\neq1$, then, although the respective inner products 
differ in the two cases, the space of functions in the two reproducing kernel Hilbert spaces is identical. 
Hereafter, we shall return to our custom of referring to the vectors and their overlap by the bra-ket 
notation so that 
 \bn  \<p'',q''|p',q'\>=(\Phi[p'',q''],\Phi[p',q'])=\K(p'',q'';p',q')\;, \en
and we let the context dictate whether or not the vectors $|p,q\>$ are normalized.

Given a reproducing kernel $\K$ depending on a number of variables as well as on various parameters, 
we can always generate alternative reproducing kernels---{\it and thereby implicitly generate their 
associated functional Hilbert spaces}---say, by rescaling variables, or by setting some of the variables to 
fixed values and thereafter ignoring them, or by suitably integrating out some of the variables, or by 
varying or even taking limits in the parameters in the reproducing kernel, etc. We shall call this general 
process a {\it reduction of the reproducing kernel}. In so doing we will invariably generate a {\it new} 
reproducing kernel---and implicitly thereby a new reproducing kernel Hibert space---but it may happen 
that the resultant reproducing kernel Hilbert space does not admit a local integral representation for its 
inner product, and even when it does it may not do so for the same measure as before or even for the same 
number of integration variables. 

At this point a few simple examples may help in understanding the reduction of reproducing kernels that 
we have in mind. Let us consider the specific example of Eq.~(21), for a single degree of freedom, which 
we denote by 
\bn \K_0(p'',q'';p',q')\equiv\exp\{i\half(p''+p')(q''-q')-\quarter[(p''-p')^2+(q''-q')^2]\}\,.  \en
 A new reproducing kernel may be obtained by rescaling the variables, for example, $p\ra\Omega^{-1} 
p$, $q\ra\Omega q$, where $0<\Omega<\infty$, which leads to
 \bn\K_1(p'',q'';p',q')=\K_0(\Omega^{-1}p'',\Omega q'';\Omega^{-1}p',\Omega q')\;.\en
 This new kernel admits a local integral representation of the inner product with the same measure and 
integration domain as before. (The result in this case is the reproducing kernel for a fiducial vector which 
is the ground state of a harmonic oscillator with angular frequency $\Omega$.) Another reproducing 
kernel arises if we simply set $p''=p'=c$, where $c$ is a constant, namely $\K_2(q'';q')=\K_0(c,q'';c,q')$. 
The result is a reproducing kernel Hilbert space that does {\it not} admit a local integral representation for 
its inner product. Other reproducing kernels arise if we integrate (45) over $p''$ and $p'$ with suitable, 
complex conjugate weight functions, e.g., $\K_3(q'';q')=\tint w^*(p'')w(p')\K_0(p'',q'';p',q')\,dp''\,dp'$. In 
addition, singular (distributional) limits may also be taken, such as the limit $\Omega\ra\infty$ for the 
first example quoted, after including a suitable $\Omega$-dependent prefactor, i.e., 
$\lim\,(\Omega/4\pi)^{1/2}\K_1(p'',q'';p',q')=\delta(q''-q')$. In such a case we no longer speak of a 
reproducing kernel Hilbert space since the new ``kernel'' cannot be expressed as the inner product of two 
normalizable and continuous vectors in a separable Hilbert space.
  
These concepts may seem somewhat abstract at this point, but we shall see specific examples of these 
kernel changes in the following sections. 
\section{Projection Operators}
Let $\E$ denote a projection operator, namely any operator that satisfies the conditions that 
$\E^{\,\dagger}=\E\,$ and $\E^{\,2}=\E\,$. The unit operator and the zero operator are projection 
operators, but our principal interest lies with projection operators that are neither ``one'' nor ``zero''. 
Given the coherent-state resolution of unity, Eq.~(20), it follows that
 \bn  \E=\E\;\one\,\E=\tint\E\,|p,q\>\<p,q|\,\E\;d\mu(p,q)  \en
holds for any projection operator $\E\,$. Taking coherent-state matrix elements of this equation leads to 
the equation
  \bn \<p'',q''|\,\E\,|p',q'\>=\tint\<p'',q''|\,\E\,|p,q\>\<p,q|\,\E\,|p',q'\>\,d\mu(p,q)\en
which along with the evident fact that \bn \<p',q'|\,\E\,|p'',q''\>=\<p'',q''|\,\E\,|p',q'\>^*\;,  \en
establishes that for each and every projection operator $\E\,$, the expression $\<p'',q''|\,\E\,|p',q'\>$ 
{\it serves as a reproducing kernel for some Hilbert space}, namely for the subspace of the original Hilbert 
space lying in the span of the projection operator $\E\,$. The dimensionality of this subspace is given, for 
example, by ${\rm Tr}(\E\,)=\tint\<p,q|\E\,|p,q\>\,d\mu(p,q)\leq\infty$.

This observation regarding projection operators will be of relevance in our analysis of constraint systems 
in that $\E\,$ will be the projection operator onto the {\it quantum constraint subspace}, and moreover, 
the kernel $\<p'',q''|\,\E\,|p',q'\>$ will serve as the reproducing kernel for the Hilbert space corresponding 
to the quantum constraint subspace, i.e., the quantum constraint Hilbert space. All the discussion 
regarding the reduction of reproducing kernels given previously applies just as well when a projection 
operator $\E\,$ is present.
\subsubsection*{Examples of projection operators} 
To fix the idea more clearly we offer a few examples of projection operators in this section. Let us first 
assume that the constraint operators $\Phi_a(P,Q)$ form a Lie algebra satisfying (14), which we initially 
assume is compact. For a compact Lie algebra, the spectrum of the self-adjoint generators is discrete, and 
they serve to generate a unitary representation of the compact Lie group the elements of which we denote 
by $\exp(-i\xi^a\Phi_a)$ for suitable real parameters $\{\xi^a\}$. Let $\delta\xi$ denote the invariant 
group measure normalized so that $\tint\delta\xi=1$ when integrated over the group manifold. Then as 
our first example of a projection operator we choose
   \bn  \E\,\equiv \tint e^{-i\xi^a\Phi_a}\,\delta\xi\;,\hskip 2cm\tint\delta\xi=1\;.  \en
It follows that 
  \bn &&\E^{\,\dagger}=\tint e^{i\xi^a\Phi_a}\,\delta\xi=\tint e^{-i\xi^a\Phi_a}\,\delta\xi^{-1}=\E\;,\\
  &&\E^{\,2}=\tint \delta\xi'\tint e^{-i(\xi'\cdot\xi)^a\Phi_a}\,\delta\xi=\E\;, \en
both equations holding in virtue of the left-equal-right invariance of the group invariant measure for 
compact groups; here $\xi^{-1}$ denotes the inverse group element to $\xi$, and ``$\cdot$'' denotes group 
composition. The projection operator determined by (50) is a projection onto the subspace for which 
$\Phi_a=0$ for all $a$. We observe in addition that 
  \bn  e^{-i\tau^a\Phi_a}\,\E\,=\E  \en
for any real $\{\tau^a\}$, which further emphasizes that $\E\,$ is a projection operator onto the zero 
subspace of the constraint operators $\{\Phi_a\}$. Finally we observe that
 \bn e^{-iT{\cal H}}\,\E\,=\E\;e^{-iT{\cal H}}=\E\;e^{-iT{\cal H}}\,\E\,\;, \en
which is an alternative statement [to (15)] of the fact that on the subspace where the constraints vanish, 
the constraints commute with the Hamiltonian.

 It is very important to record that {\it the integral over the group elements is actually an average}, as is 
clear from the normalization $\tint\,\delta\xi=1$.

We next consider a noncompact group some of whose generators have a continuous spectrum. For a 
noncompact group the integral of the invariant measure over the group manifold diverges, and so we 
define our projection operator to be
 \bn  \E\,=\tint e^{-i\xi^a\Phi_a}\,f(\xi)\,\delta\xi\;,  \en
where in the present case $\delta\xi$ denotes the left-invariant group measure. In order that this 
expression represents a projection operator it is necessary and sufficient that $f^*(\xi^{-1})\,\delta
\xi^{-1}=f(\xi)\,\delta\xi$, so as to satisfy $\E^{\,\dagger}=\E\,$, and 
\bn  \tint f(\xi')f(\xi'^{-1}\cdot\xi)\,\delta\xi'=f(\xi)\;, \en
 so as to satisfy $\E^{\,2}=\E\,$. It follows from the second condition that $(\tint f(\xi)\,\delta\xi)^2\!=\tint 
f(\xi)\,\delta\xi$, which for the case at hand means that $\tint f(\xi)\,\delta\xi=1$ representing an average 
over the group just as was true for a compact group.

For constraint-operator generators that form a Lie algebra and yet have a continuous spectrum, it follows 
that we cannot project exclusively onto the spectral value zero, reflecting the same problem that we stated 
earlier arose in the Dirac formulation. But we can project onto a small interval around zero. In particular, 
if $\delta>0$ denotes a small positive parameter (e.g., $\delta=10^{-50}$), then we can arrange to 
construct a projection operator $\E\,$ such that for all $|\psi\>$ for which $\E\,|\psi\>=|\psi\>$ we have 
$\<\psi|\Sigma_a\Phi^2_a|\psi\>\leq\delta^2\<\psi|\psi\>$.

Let us consider the example of a single momentum operator $P$ and let the single parameter $\xi\in\ir$. 
Then we choose
  \bn \E\,\equiv \int_{-\infty}^\infty e^{-i\xi P}\,\frac{\sin(\delta\xi)}{\pi\xi}\,d\xi =\E\,(-\delta<P<\delta)\;;  
\en namely, just as the notation indicates, the integral results in a projection operator for the operator $P$ 
onto just the interval $(-\delta,\delta)$. The new reproducing kernel in this case is given by
 \bn &&\hskip-1.2cm\<p'',q''|\E\,|p',q'\>=\pi^{-1/2}\int_{-\delta}^{\delta}\exp[-\half(k-p'')^2+ik(q''-q')-
\half(k-p')^2]\,dk\nonumber\\&&\hskip1.5cm=\frac{2\sin[\delta(q''-q')]}{\sqrt{\pi}(q''-q')}\,\exp
[-\half(p''^2+p'^2)]+O(\delta^2)\;,  \en
where the second line is an approximate evaluation accurate, as indicated, to leading order in $\delta$. 
Note that, by itself, the leading term of order $\delta$ represents a reproducing kernel which has a local 
integral representation for the inner product with the same measure and integration domain as the 
reproducing kernel $\<p'',q''|p',q'\>$ without the projection operator $\E\,$.
\section{Application to First-Class Constraints}
\subsubsection*{Compact groups}
Let us consider the coherent-state expression appropriate to the propagator
 \bn \<p'',q''|{\sf T}e^{-i\tint[{\cal H}+\l^a(t)\Phi_a]\,dt}\,|p',q'\>\;. \en
Here $\sf T$ denotes the time-ordering operation, and at this point $\{\l^a(t)\}$ just represent some 
external, time-dependent functions. Based on the discussion of Sec.~2, it follows that this propagator 
admits the formal path integral
 \bn \int\exp\{i\tint[p_j{\dot q}^j-H(p,q)-\l^a\phi_a(p,q)]\,dt\}\,\D\mu(p,q)\;,\en
which evidently depends on whatever choice may be made for the functions $\{\l^a(t)\}$. 

Now let us impose the quantum analog of the initial value equation. For the present we assume that the 
constraint operators $\Phi_a$ and the Hamiltonian together satisfy (14) and (15), and therefore the 
constraints constitute a Lie algebra which for the present we also assume relates to a compact group. Let 
  \bn \E\,\equiv\tint e^{-i\xi^a\Phi_a}\,\delta\xi\;,\hskip2cm\tint\delta\xi=1  \en
represent the projection operator onto the subspace for which $\Phi_a=0$ for all $a$, which defines the 
quantum constraint subspace. We impose this subspace condition on the propagator (59) through the 
integral
\bn\int\<p'',q''|{\sf T}e^{-i\tint[{\cal H}+\l^a(t)\Phi_a]\,dt}\,|{\o p}',{\o q}'\>\<{\o p}',
{\o q}'|\E\,|p',q'\>\,d\mu({\o p}',{\o q}')\;. \en 
On the one hand, this expression has the formal path-integral representation given by
 \bn \int\exp\{i\tint[p_j{\dot q}^j-H(p,q)-\l^a\phi_a(p,q)]\,dt\}\,
e^{-i\xi^a\phi_a(p',q')}\,\D\mu(p,q)\,\delta\xi\;,  \en
while on the other hand, it follows that
 \bn &&\hskip-1cm\int\<p'',q''|{\sf T}e^{-i\tint[{\cal H}+\l^a(t)\Phi_a]\,dt}\,|{\o p}',{\o q}'\>\<{\o p}',
{\o q}'|\E\,|p',q'\>\,d\mu({\o p}',{\o q}')\nonumber\\
&&=\<p'',q''|{\sf T}e^{-i\tint[{\cal H}+\l^a(t)\Phi_a]\,dt}\,\E\,|p',q'\>\nonumber\\
&&=\lim\,\<p'',q''|[\prod^{\leftarrow}_l(e^{-i\epsilon{\cal H}}
e^{-i\epsilon\l^a_l\Phi_a})]\,\E\,|p',q'\>\nonumber\\
&&=\<p'',q''|e^{-iT{\cal H}}e^{-i\tau^a\Phi_a}\,\E\,|p',q'\>\nonumber\\
&&=\<p'',q''|e^{-iT{\cal H}}\,\E\,|p',q'\>\;.  \en
In these relations we have employed: (i) the resolution of unity to bring the projection operator inside the 
propagator, (ii) a time-ordered lattice approximation together with the Trotter product formula, (iii) 
Eqs.~(14) and (15) to bring all constraint group operators together where $\{\tau^a\}$ are constructed 
from the functions $\{\l^a\}$, and lastly (iv) Eq.~(53) to eliminate any dependence on $\{\tau^a\}$ and 
therefore to eliminate any dependence on the functions $\{\l^a\}$. To summarize, we have established a 
definition of the coherent-state path integral for which 
\bn &&\hskip-1cm\int\exp\{i\tint[p_j{\dot q}^j-H(p,q)-\l^a\phi_a(p,q)]\,dt-
i\xi^a\phi_a(p',q')\}\,\D\mu(p,q)\,\delta\xi\nonumber\\
&&\hskip.1cm=\<p'',q''|e^{-iT{\cal H}}\,\E\,|p',q'\>\;.  \en
Observe carefully what this relation shows: On the left-hand side there is a path integral that apparently 
depends on the fixed but arbitrary time-dependent functions $\{\l^a(t)\}$, $0<t<T$; however, the 
right-hand side demonstrates that {\it in fact the path integral is completely independent of the functions} $\{\l^a\}\,!$

If the left-hand side is completely independent of the functions $\{\l^a(t)\}$, then we are free to {\it 
average} the left-hand side over the functions $\l^a$ with a general complex measure we denote by 
$C(\l)$ for which $\tint \D C(\l)\equiv 1$ and for which we only require that such an average should 
introduce (at least) one projection operator factor $\E\,$.
Thus we also conclude that
\bn &&\hskip-1cm\int\exp\{i\tint[p_j{\dot q}^j-H(p,q)-\l^a\phi_a(p,q)]\,dt\}\,\D\mu(p,q)\,
\D C(\l)\nonumber\\
&&\hskip.1cm=\<p'',q''|e^{-iT{\cal H}}\,\E\,|p',q'\>\;,  \en 
provided $\tint\D C(\l)=1$ and that such an average over the functions $\{\l^a\}$ introduces a factor 
$\E\,$. At this point, on the left-hand side, we recognize the path integral appropriate to a system with 
first-class constraints in which the dynamical degrees of freedom $(p,q)$ are integrated with the canonical 
measure $\D\mu(p,q)$ and the auxiliary variables $\l^a$ are to be averaged with respect to a (possibly 
complex) measure $\D C(\l)$, $\tint\D C(\l)=1$. 
\subsubsection*{Commentary}
From the point of view of the classical theory this prescription leads to the classical action
  \bn I=\tint[p_j{\dot q}^j-H(p,q)-\l^a\phi_a(p,q)]\,dt  \en
the equations of motion of which arise as usual from a stationary variation principle of the variables 
involved, namely, variations with respect to $p_j$, $q^j$, and $\l^a$. At this point it becomes clear that 
the variables $\{\l^a\}$ are to be interpreted classically as Lagrange multipliers and not as conventional 
dynamical degrees of freedom. 

Finally, we emphasize once again, that with the proper measure accorded to the Lagrange multipliers, the 
result of the path integration is completely independent of the choice of the measure $C(\l)$ provided only 
that it is normalized and does in fact implement the quantum ``initial value equation'' in the sense of a 
projection onto the quantum constraint subspace. No $\delta$-functionals have arisen, and indeed could 
not arise from an integration over Lagrange multipliers that is an {\it average}. With no 
$\delta$-functionals of the constraints appearing, there are no divergent integrals that require gauge 
fixing, and 
finally, therefore, no Faddeev-Popov determinant is introduced. The quantum constraint subspace is {\it 
already gauge invariant} as demonstrated by the independence of (65) on the (now recognized to be) 
gauge parameters $\{\l^a\}$. Observables, which satisfy the quantum analog of (8), are (just like the 
Hamiltonian) gauge invariant in the quantum constraint subspace. It is not that the quantum dynamics 
needs to specify  a choice of the gauge variables---as the classical dynamics certainly does---but rather the 
quantum theory, properly formulated, is {\it already independent of these unphysical variables!} All these 
positive features have emerged simply by paying due respect to what measure should be taken for the 
Lagrange multipliers in order to realize the quantum constraint subspace!

Examples of such a quantization procedure are reserved to Sec.~6.
\subsubsection*{Noncompact groups}
We next develop a parallel analysis in the case of noncompact groups, and particularly for those cases 
where some of the constraint operators $\Phi_a$ have a continuous spectrum. As argued in Sec.~4 we 
cannot in this case choose the same expression for the projection operator because the group volume is 
infinite. Moreover, we cannot project onto the value $\Phi_a=0$ for operators with a continuous spectrum 
but instead we will settle (initially) for a projection operator $\E\,$ that projects onto the subspace for 
which $\Sigma_a\Phi^2_a\leq\delta^2$ for some arbitrarily small parameter $\delta>0$. (At the 
conclusion of this section, as well as in an example in Sec.~6, we discuss the limit $\delta\ra0$.) With 
such qualifying remarks, the discussion in the case of noncompact groups is very similar to that already 
given for compact groups.

Let us start directly with the expression
 \bn \<p'',q''|{\sf T}e^{-\tint[{\cal H}+\l^a(t)\Phi_a]\,dt}\,\E\,|p',q'\> \en
which we have arrived at by imposing the quantum ``initial value equation'', just as in the case of compact 
groups, except in the present case $\E\,$ is to be defined by Eq.~(55). The same manipulations that 
applied in the compact group case also bring us to the relation
 \bn \<p'',q''|e^{-iT{\cal H}}e^{-i\tau^a\Phi_a}\,\E\,|p',q'\>\;, \en
but now there is no direct analog of (53) which would permit us to drop completely the term involving 
$\tau^a$ since the operators $\Phi_a$ are not zero on the space spanned by $\E\,$. While the constraint 
operators are not zero on this space, they are nevertheless very small, namely of order $\delta$, which is 
as small as we choose. In particular, {\it the $\tau$ dependence appears only in higher-order corrections to 
the leading order in $\delta$ that is represented by} (66); for $\delta=10^{-50}$ such corrections would be 
negligible. Thus, we conclude, {\it to leading order in} $\delta$, the dependence on $\tau$, and therefore 
on the original functions $\l^a(t)$ disappears from the path integral expression, and we are led to an 
expression similar to that found for compact groups, namely
 \bn &&\hskip-1cm\int\exp\{i\tint[p_j{\dot q}^j-H(p,q)-\l^a\phi_a(p,q)]\,dt\}\,\D\mu(p,q)\,
\D C(\l)\nonumber\\
&&=\<p'',q''|e^{-iT{\cal H}}\,\E\,|p',q'\>[1+O(\delta)]\;, \en
where $C(\l)$ is a (possibly complex) normalized measure, i.e., $\tint\D C(\l)=1$, which is chosen so that 
at least one projection operator $\E\,$ appears within the propagator expression. All other dependence of 
the path integral on the measure $C(\l)$, or stated otherwise, all dependence on the gauge variables is 
contained in the higher-order terms represented by the factor $O(\delta)$ in (70). A similar conclusion is 
obtained regarding any observable in the present case, namely, that any gauge dependence would lie in 
higher-order corrections in $\delta$.
\subsubsection*{Limit $\delta\ra0$}
Although when $\delta$ is extremely tiny corrections to proper answers are negligibly small, one can also 
pass to the limit $\delta\ra0$, and
a few words on how that limit can be taken are in order. We first observe, in light of (54), that
 \bn \<p'',q''|\E\,e^{-iT{\cal H}}\,\E\,|p',q'\>=\<p'',q''|e^{-iT{\cal H}}\,\E\,|p',q'\>\;.  \en
Although this expression was established in Sec.~3 for a compact group, it holds in the case of 
noncompact groups to leading order in $\delta$.
 Consequently, the expression 
  \bn \K(p'',q'',t'';p',q',t')\equiv\<p'',q''|\E\,e^{-i(t''-t'){\cal H}}\,\E\,|p',q'\>\en
defines a reproducing kernel in the sense of Sec.~3. As such we may consider a convergent limit that 
leads to a bounded and continuous function, which will also correspond to a reproducing kernel generally 
different than the one we started with. In particular, suppose, as a consequence of the choice of $\E\,$, that 
the path integral represented by (70) has a leading order of $\delta^\sigma$. In that case, we introduce the kernel
 \bn \K_1(p'',q'',t'';p',q',t')\equiv\lim_{\delta\ra0}\,\delta^{-\sigma}\K(p'',q'',t'';p',q',t')  \en
which defines a reproducing kernel for the true quantum constraint subspace for which ``$\Phi_a=0$''. In 
this way all effects that are higher order in $\delta$ are eliminated, and a description of the quantum 
system that is fully as gauge invariant as was the case for a compact group is attained for the case of a 
noncompact group with some generators having continuous spectrum. As discussed in Sec.~3, although 
the limiting expression generally leads to a reproducing kernel, which therefore characterizes its 
associated reproducing kernel Hilbert space, it is not clear in general whether or not the inner product of 
the resultant reproducing kernel Hilbert space admits a local integral representation with a positive 
measure; that must be decided on a case by case study. 

A simple illustration of what may happen in the limit $\delta\ra0$ can be given by examining the example 
in (58). In particular, consider the limit
 \bn &&\hskip-1cm\K(p'',q'';p',q')\equiv\lim_{\delta\ra0}(\sqrt{\pi}/2\delta)\{\frac{2\sin[\delta
(q''-q')]}{\sqrt{\pi}(q''-q')}\,\exp[-\half(p''^2+p'^2)]+O(\delta^2)\}\nonumber\\
&&\hskip1.7cm=\exp[-\half(p''^2+p'^2)]\;.  \en
The resultant function is bounded and continuous, and it defines a reproducing kernel, but one that is  
independent of the variables $q''$ and $q'$. This independence implies that we have at last arrived at the 
desired quantum constraint subspace for this example, one where ``$P=0$''. The resultant functional 
Hilbert space is {\it one dimensional}, a perfectly acceptable result, and every vector  in the reproducing 
kernel Hilbert space is proportional to $\exp(-p^2/2)$. In the present case, the reproducing kernel Hilbert 
space admits a local integral representation for the inner product based on the measure $dp/\sqrt{\pi}$ 
integrated over the real line. Of course, it is a bit of an overkill to use such heavy machinery for a one-
dimensional Hilbert space and alternative, but equivalent, characterizations are much simpler. 
Nevertheless, this example serves to illustrate how a reduction of the reproducing kernel may take place 
and how the dimensionality of the Hilbert space can change dramatically.

\section{Examples of First-Class Constraints}
\subsubsection*{General configuration space geometry}
Although we shall discuss constraints that lead to a general configuration space geometry in this section, 
we shall for the most part use rather simple illustrative examples. To begin with let us consider the 
constraint
 \bn   \sum_{j=1}^J(q^j)^2=1\;,  \en
a condition which puts the classical dynamics on a (hyper)sphere of unit radius. For convenience in what 
follows we shall focus as well on the case of a vanishing Hamiltonian so as to isolate clearly the 
consequences of the constraint independently of any dynamical effects (for a discussion of compatible 
dynamics see \cite{ikl}---and for {\it non}compatible dynamics see Sec.~8). Employing a standard vector 
inner product notation, we consider the formal path integral
  \bn {\cal M}\int\exp\{i\tint[p\cdot{\dot q}-\l(q^2-1)]\,dt\}\,\D p\,\D q\,\D C(\l)\;,  \en
the result of which in the light of the discussion in Sec.~5, is given by
  \bn  \<p'',q''|\E\,|p',q'\> \en
where
  \bn  \E&&=\int_{-\infty}^\infty e^{-i\l(Q^2-1)}\,\frac{\sin(\delta\l)}{\pi\l}\,
d\l    =\E\,(-\delta<Q^2-1<\delta)\;.  \en

In order ultimately to obtain a suitable (reduced) reproducing kernel in the present case we  allow for 
fiducial vectors other than harmonic oscillator ground states. Thus we let $|\eta\>$ denote a general unit 
vector for the moment; its required properties will emerge from our analysis. We choose a phase 
convention for the coherent states---in particular, in (19) we set $\alpha(p,q)=pq$, rather than zero---so 
that the Schr\"odinger representation of the coherent states reads
 \bn \<x|p,q\>=e^{ip\cdot x}\,\eta(x-q)\;,  \en
which leads immediately to the expression
  \bn \<p'',q''|p',q'\>=\int\eta^*(x-q'')\,e^{-i(p''-p')\cdot x}\,\eta(x-q')\,d^J\!x\;. \en
Consequently, the reproducing kernel that incorporates the projection operator is given by
  \bn \<p'',q''|\E\,|p',q'\>=\int_{1-\delta<x^2<1+\delta}\eta^*(x-q'')\,e^{-i(p''-p')\cdot x}\,\eta(x-q')\,d^J\!x\;. 
\en
Since $\E\,$ represents a projection operator, it is evident that this expression defines a reproducing kernel 
that admits a local integral for its inner product (for any normalized $\eta$) with a measure 
$d^J\!p\,d^J\!q/(2\pi)^J$ and an integration domain $\ir^{2J}$.

However, if we are willing to restrict our choice of fiducial vector, we can reduce the number of 
integration variables and change the domain of integration in a meaningful way. Recall that the group 
${\rm E}(J)$, the Euclidean group in $J$-dimensions, consists of rotations that preserve the unit 
(hyper)sphere in $J$-dimensions, as well as $J$ translations. As emphasized by Isham \cite{ish} this is 
the appropriate ``phase space'' for a system confined to the surface of a (hyper)sphere in $J$ dimensions. 
We can adapt our present coherent states to be coherent states for the group ${\rm E}(J)$ without 
difficulty.

To that end consider the reduction of the reproducing kernel (81) to one for which $q''^2=q'^2\equiv1$. 
To illustrate the process as clearly as possible let us choose $J=2$. As a consequence we introduce
  \bn  \<a'',b'',c''|a',b',c'\>\equiv\<p'',q''|\E\,|p',q'\>_{q''^2=q'^2=1}\;, \en
where $a\equiv p^1$, $b\equiv p^2$, and $c$ arises from the identification $q^1\equiv\cos(c)$ and
 $q^2\equiv\sin(c)$, all relations holding for both end points. Expressed in terms of polar coordinates, 
$r,\phi$, the reduced reproducing kernel becomes
  \bn &&\hskip-1.2cm\<a'',b'',c''|a',b',c'\> \nonumber \\   &&\hskip-.7cm=\int_{|r^2-1|<\delta}\eta^*(r,\phi-
c'')\,e^{-i(a''-a')r\cos\phi-i(b''-b')r\sin\phi}\,\eta(r,\phi-c')\,r\,dr\,d\phi\,.  \en

We next seek to choose $\eta$, if at all possible, in such a way that the inner product of this new (reduced) 
reproducing kernel admits a local integral for its inner product. As a starting point we choose the left-
invariant group measure for ${\rm E}(2)$ which is given by $M\,da\,db\,dc$, $M$ a constant, with an 
integration domain $\ir^2\times S^1$. Therefore, we are led to study \bn&&\hskip-1cm\int
\int_{|r^2-1|<\delta}\eta^*(r,\phi-c'')\,e^{-i(a''-a)r\cos\phi-i(b''-b)r\sin\phi}\,\eta(r,
\phi-c)\,r\,dr\,d\phi\nonumber\\
&&\hskip-.1cm\times\int_{|\rho^2-1|<\delta}\eta^*(\rho,\theta-c)\,e^{-i(a-a')\rho\cos\theta-
i(b-b')\rho\sin\theta}\,\eta(\rho,\theta-c')\,\rho\,d\rho\,d\theta\nonumber\\
&&\hskip-.1cm\times M\,da\,db\,dc  \nonumber\\
&&\hskip-.5cm=(2\pi)^2M\int\eta^*(r,\phi-c'')e^{-i(a''-a')r\cos\phi-i(b''-b')r\sin\phi}\,
\eta(r,\phi-c')\,r\,dr\,d\phi\nonumber\\
&&\hskip-.1cm\times\int|\eta(r,c)|^2\,dc\;,  \en
which leads to the desired result provided (i)
  \bn  \int_0^{2\pi}|\eta(r,c)|^2\,dc=P\;,\hskip1cm P>0\;,  \en
is {\it independent} of $r$, $|r^2-1|<\delta$, and (ii) $M=[(2\pi)^2\,P]^{-1}$. Given a general 
nonvanishing vector $\xi(r,\phi)$ a vector satisfying (85) may always be given by
  \bn  \eta(r,\phi)=\xi(r,\phi)/\sqrt{\tint_0^{2\pi}|\xi(r,\theta)|^2\,d\theta} \en   provided the denominator is 
positive, and 
which therefore leads to $P=1$. In this way we have reproduced the ${\rm E}(2)$-coherent states of 
Ref. ~\cite{ikl}, even including the necessity for a small interval of integration in $r$, and where fiducial 
vectors satisfying (85) were called ``surface constant''.

Dynamics consistent with the constraint $q^2=1$ is obtained in the ${\rm E}(2)$ case by choosing a 
Hamiltonian that is a function of the coordinates on the circle, namely $\cos(\theta)$ and $\sin(\theta)$, as 
well as the rotation generator of ${\rm E}(2)$, i.e., $-i\d/\d\theta$. We refer the reader to \cite{ikl} for a 
further discussion of ${\rm E}(2)$-coherent states as well as a discussion of the introduction of compatible 
dynamics. An analogous discussion can be given for the classical constraint $q^2=1$ for any value of 
$J>2$.

Not only can compact (hyper)spherical configuration spaces be treated in this way, but we may also treat 
noncompact (hyper)pseudospherical spaces defined by the constraint
  \bn \Sigma_{i=1}^Iq^{i\,2}-\Sigma_{j=I+1}^Jq^{j\,2}=1\;,\hskip1cm1\leq I\leq J-1\;,  \en
appropriate to the Euclidean group ${\rm E}(I,J-I)$. Such an analysis leads to ${\rm E}(I,J-I)$-coherent 
states. 

Finally, we comment on the constraint of a general curved configuration space which can be defined by a 
set of compatible constraints $\phi_a(q)=0$.
Clearly these constraints satisfy $\{\phi_a(q),\phi_b(q)\}=0$, and define a $(J-A)$-dimensional 
configuration space in the original Euclidean configuration space $\ir^J$. The relevant projection operator 
$\E\,=\E\,(\Sigma\Phi_a^2(Q)<\delta^2)$ is defined in an evident fashion, and the reproducing kernel 
incorporating the projection operator is defined in analogy with the prior discussion. This reproducing 
kernel enjoys a local integral representation for its inner product, in fact, it is with the same measure and 
integration domain as without the projection operator. What differs in the present case is that when the 
reproducing kernel is put on the constraint manifold, the resultant coherent states are generally {\it not} 
defined by the action of a group on a fixed fiducial vector. In short, the relevant coherent states are not 
group generated. Such a situation is not unknown \cite{kl12}. 

We defer a discussion of dynamics in the present case until Sec.~8.
\subsubsection*{Finite-dimensional Hilbert spaces}
Let us consider the case of two degrees of freedom with a ``classical'' action function given by
\bn I =\tint[\half(p_1{\dot q}_1-q_1{\dot p}_1+p_2{\dot q}_2-q_2{\dot p}_2)-
\l(p_1^2+p_2^2+q_1^2+q_2^2-4s\hbar)]\,dt  \en
Note that for notational convenience we have departed from our conventional index placement for 
coordinates, that we have explicitly included $\hbar$ in our ``classical'' action, and that we have 
effectively chosen another phase convention for the associated coherent states [equivalent to (19) with 
$\alpha(p,q)=pq/2$]. With this alternative phase convention the unconstrained reproducing kernel is 
given by
 \bn &&\<p'',q''|p',q'\>\equiv\<z''|z'\> \nonumber\\
&&\hskip2.3cm=\exp[\Sigma_{j=1}^2(-\half|z''_j|^2+{z''^*}_jz'_j-\half|z'_j|^2)]  \en
where $z_j\equiv (q_j+ip_j)/\sqrt{2\hbar}$ for each of the end points.

We next observe that the operator constraint equation
 \bn \Phi=:P^2_1+P^2_2+Q^2_1+Q^2_2:-4s\hbar\one  \en
has discrete eigenvalues, i.e., $2(n_1+n_2-2s)\hbar$, where $n_1$ and $n_2$ are nonnegative integers, 
based on the choice of $|\eta\>$ as the ground state of each oscillator. To satisfy $\Phi=0$ it is necessary 
that $2s$ be an integer in which case the quantum constraint subspace is $(2s+1)$-dimensional. The 
projection operator in the present case is defined by
 \bn  \E\,=\pi^{-1}\int_0^{\pi} \exp[\,i\l(:P^2_1+P^2_2+Q^2_1+Q^2_2:-4s\hbar\one)/\hbar]\,d\l\;,  \en
and projects onto a $(2s+1)$-dimensional subspace. It is straightforward to demonstrate that
  \bn &&\hskip-1.3cm\<z''|\E\,|z'\>=\exp[-\half\Sigma_{j=1}^2(|z''_j|^2+|z'_j|^2)] 
{[(2s)!]}^{-1}({z''^*}_1z'_1+{z''^*}_2z'_2)^{2s}\nonumber\\
&&\hskip-1cm=\exp[-\half\Sigma_{j=1}^2(|z''_j|^2+|z'_j|^2)] \sum_{k=0}^{2s}
[{k!(2s-k)!}]^{-1}({z''^*}_1z'_1)^k({z''^*}_2z'_2)^{2s-k}\,.   \en
The projected reproducing kernel in this case corresponds to a {\it finite} dimensional Hilbert space; 
nevertheless, the inner product is given by the same measure and integration domain as in the original 
unprojected, infinite dimensional Hilbert space!

As the notation suggests the present quantum constraint subspace provides a natural carrier space for an 
irreducible representation of $\rm SU(2)$ of spin $s$.  We observe that the following three expressions 
represent generators of the rotation group in their action on the constraint hypersurface:
 \bn &&S_x=\half(p_1p_2+q_1q_2)\;,  \nonumber\\
     &&S_y=\half(q_1p_2-p_1q_2)\;,  \nonumber\\
     &&S_z=\quarter(p_1^2+q_1^2-p^2_2-q_2^2)\;.   \en
Thus these quantities serve as potential ingredients to a Hamiltonian which is compatible with the 
constraint.

Of course, there are other, simpler and more familiar ways to represent a finite-dimensional Hilbert space; 
but any other representation is evidently equivalent to the one described here. 

Although not the subject of this section, we may also point out that an analogous discussion holds in case 
of the constraint
\bn \phi(p,q)=p_1^2+q_1^2-p_2^2-q_2^2-4k\hbar=0\;,   \en
where $k$ is an integer, and the resultant reduced Hilbert space is infinite dimensional for any integral 
$k$ value.
\subsubsection*{FLPR model}
In a recent paper \cite{lee}, Friedberg, Lee, Pang, and Ren introduced a model sensitive to the problem of 
Gribov ambiguities. (For the details of the model and its possible role as a simple analogue of non-Abelian 
gauge models, we refer the reader to their paper.) We begin with the classical Hamiltonian for a three-
degree of freedom system given by
\bn H=\half(p_1^2+p_2^2+p_3^2)+U(q_1^2+q_2^2)+\l[g(p_2q_1-q_2p_1)+p_3]\;,\en
where $U$ denotes the potential, which hereafter, following \cite{lee}, we shall choose as harmonic, 
namely $U(q_1^2+q_2^2)=\omega^2(q_1^2+q_2^2)/2$, because then this special model is fully soluble. 
Here, $g>0$ is a coupling constant, and $\l=\l(t)$ is the Lagrange multiplier which enforces the single 
first-class constraint
 \bn \phi(p,q)=g(p_2q_1-q_2p_1)+p_3=0\;.  \en

For the first two degrees of freedom we choose coherent states with the phase convention adopted for the 
previous example, while for the third degree of freedom we maintain the usual phase convention. This 
choice means that we consider the formal coherent state path integral given by
  \bn &&\hskip-1cm\int\exp\{i\tint[\half(p_1{\dot q}_1-q_1{\dot p}_1)+\half(p_2{\dot q}_2-
q_2{\dot p}_2)+p_3{\dot q}_3\nonumber\\&&\hskip.8cm-\half(p_1^2+p_2^2+p_3^2)-
\half\omega^2(q_1^2+q_2^2)\nonumber\\&&\hskip.8cm+\l[g(p_2q_1-q_2-
p_1)+p_3]\,dt\}\,\D\mu(p,q)\,\D C(\l)\nonumber\\
   &&=\<z''_1,z''_2,p''_3,q''_3|\,e^{-i{\cal H}T}\,\E\,|z'_1,z'_2,p'_3,q'_3\>\;.  \en
In the present case the relevant projection operator $\E\,$ is given (for $\hbar=1$, and $0<\delta\ll g$) by
 \bn \E\,=\E\,(-\delta<gL_3+P_3<\delta)=\sum_{m=-\infty}^\infty\E\,
(-\delta<gm+P_3<\delta)\,\E\,(L_3=m)\;,  \en
where we have used the familiar spectrum for the rotation generator $L_3$.
If ${\cal H}_0$ denotes the harmonic oscillator Hamiltonian for the first two degrees of freedom, then it 
follows that 
\bn &&\hskip-.3cm\<z''_1,z''_2,p''_3,q''_3|\,e^{-i{\cal H}T}\,\E\,|z'_1,z'_2,p'_3,q'_3\>\nonumber\\
 &&=\sum_{m=-\infty}^\infty\<z''_1,z''_2|e^{-i{\cal H}_0T}\E\,(L_3=m)\,|z'_1,z'_2\>\nonumber\\  &&\hskip.3cm\times\<p''_3,q''_3|e^{-iP_3^2T/2}\E(-\delta<gm+P_3<\delta)|p'_3,q'_3\>\nonumber\\
&&=\exp[-\half(|z''_1|^2+|z''_2|^2+|z'_1|^2+|z'_2|^2)]\nonumber\\
&&\hskip.3cm\times\sum_{m=-\infty}^\infty\Big\{\frac{(z''^*_1+iz''^*_2)(z'_1-iz'_2)}{(z''^*_1-
iz''^*_2)(z'_1+iz'_2)}\Big\}^{m/2}\,I_m(\sqrt{(z''^{*2}_1+z''^{*2}_2)(z'^2_1+z'^2_2)}e^{-i\omega 
T})\nonumber\\
&&\hskip.3cm\times\exp[-\half(gm+p''_3)^2-\half(gm+p'_3)^2-i\half g^2m^2T-igm(q''_3-
q'_3)]\nonumber\\
&&\hskip.3cm\times\frac{2}{\sqrt{\pi}}\frac{\sin[\delta(q''_3-q'_3)]}{(q''_3-q'_3)}+O(\delta^2)\;,  \en
where $I_m$ denotes the usual Bessel function.

We observe that the spectrum for the Hamiltonian agrees with the results of Ref. \cite{lee}, and moreover, 
that we have obtained gauge-invariant results, i.e., insensitivity to any choice of the Lagrange multiplier 
function $\l(t)$, merely by projecting onto the quantum constraint subspace at $t=0$. The constrained 
propagator (99) is composed with the same measure and integration domain as is the unconstrained 
propagator. Just as was the case in the first example in this section, we may also divide the constrained 
propagator by $\delta$ and take the limit $\delta\ra0$. The result is a new functional expression for the 
propagator that fully satisfies the constraint condition, but no longer admits an inner product with the 
same measure and integration domain as before. 

\section{Application to General Constraints}
\subsubsection*{Classical considerations}
When dealing with a general constraint situation it will typically happen that the self-consistency of the 
equations of motion will determine some or all of the Lagrange multipliers in order for the system to 
remain on the classical constraint hypersurface. If the Poisson brackets of the constraints themselves do 
not all vanish on the constraint hypersurface then the Lagrange multipliers assume given values so that 
the constraints are maintained. In addition, if the Hamiltonian attempts to force points initially lying on 
the constraint hypersurface to leave that hypersurface, then it is the task of the Lagrange multipliers to 
supply the necessary forces for the system to remain on the constraint hypersurface. This standard 
interpretation of what happens at the classical level provides the clue as to how to proceed at the quantum 
level!
\subsubsection*{Quantum considerations}
As in previous sections we let $\E\,$ denote the projection operator onto the quantum constraint subspace. 
On the basis of the classical discussion given above we are motivated to consider the quantity
\bn \lim\,\<p'',q''|\E\,e^{-i\epsilon{\cal H}}\E\,e^{-i\epsilon{\cal H}}\cdots\E\,e^{-i\epsilon
{\cal H}}\E\,|p',q'\> \en
where the limit, as usual, is for $\epsilon\ra0$. The physics behind this expression is as follows. Reading 
from right to left we first impose the quantum initial value equation, and then propagate for a small 
amount of time ($\epsilon$). Next we recognize that the system may have left the quantum constraint 
subspace, and so we project it back onto that subspace, and so on over and over. In the limit that 
$\epsilon\ra0$ the system remains within the quantum constraint subspace and (100) actually leads to
 \bn \<p'',q''|\E\,e^{-iT(\E\;{\cal H}\E\;)}\E\,|p',q'\>\;, \en
an expression that clearly illustrates temporal evolution entirely within the quantum constraint subspace. 
In fact the situation is not as simple as that remark would seem to make it. Although we assume that the 
original Hamiltonian $\cal H$ is self adjoint, and thus generates a unitary time evolution in the original 
Hilbert space, the operator $\E\,{\cal H}\E\,$ may either (i) be (essentially) self adjoint, (ii) admit self-
adjoint extensions, or (iii) admit no self-adjoint extension at all. Although examples of all three 
possibilities are easily given, we shall assume for the sake of convenience, that the resultant operator 
$\E\,{\cal H}\E\,$ is itself a self-adjoint operator (or the closure of an essentially self-adjoint operator 
which we denote by the same symbol). With that assumption we conclude that (101) describes a unitary 
time evolution within the quantum constraint subspace. 

The two equivalent expressions given in the preceding paragraph may be developed in two other ways. 
First, we repeatedly insert the coherent-state resolution of unity in such a way that (100) becomes
 \bn \lim\,\int\prod_{l=0}^N\<p_{l+1},q_{l+1}|\E\,e^{-i\epsilon
{\cal H}}\E\,|p_l,q_l\>\prod_{l=1}^Nd\mu(p_l,q_l)\;. \en
We wish to turn this expression into a formal path integral, but the procedure used previously relied on 
the use of unit vectors. Although the coherent-state vectors $|p,q\>$ are unit vectors, there is no guarantee 
that the vectors $\E\,|p,q\>$ are unit vectors. Thus let us rescale the factors in the integrand introducing 
  \bn  |p,q\>\!\>\equiv\E\,|p,q\>/\|\E\,|p,q\>\|  \en
which are unit vectors. If we let $M''=\|\E\,|p'',q''\>\|$, $M'=\|\E\,|p',q'\>\|$, and observe that 
$\|\E\,|p,q\>\|^2=\<p,q|\E\,|p,q\>$, it follows that (102) may be rewritten as
 \bn M''M'\lim\,\int\prod_{l=0}^N\<\!\<p_{l+1},q_{l+1}|e^{-i\epsilon
{\cal H}}|p_l,q_l\>\!\>\prod_{l=1}^N\<p_l,q_l|\E\,|p_l,q_l\>\,d\mu(p_l,q_l)\;. \en
In turn this expression is represented by the formal path integral
\bn M''M'\int\exp\{i\tint[i\<\!\<p,q|\frac{d}{dt}|p,q\>\!\>-\<\!\<p,q|
{\cal H}|p,q\>\!\>]\,dt\}\,\D_E\mu(p,q)\;,  \en
where the new formal measure for the path integral is defined in an obvious fashion from its lattice 
prescription. We can also reexpress this formal path integral in terms of the original bra and ket vectors as 
 \bn &&\hskip-1cm M''M'\int\exp\{i\tint[i\<p,q|\E\,\frac{d}{dt}|p,q\>/\<p,q|\E\,|p,q\>\nonumber\\
    &&\hskip2.3cm-\<p,q|\E\,{\cal H}\E\,|p,q\>/\<p,q|\E\,|p,q\>]\,dt\}\,\D_E\mu(p,q)\;.\en
This last relation is the end of our second route of calculation beginning with (100).

The third relation we wish to derive uses the integral representation for the projection operator $\E\,$. In 
this analysis we assume that $\epsilon\equiv T/N$ [rather than our customary $\epsilon=T/(N+1)$]. Thus 
we rewrite (100) in the form
\bn &&\hskip-1.5cm\lim\int\<p'',q''|e^{-i\epsilon\l^a_{N}\Phi_a}e^{-i\epsilon{\cal H}}
e^{-i\epsilon\l^a_{N-1}\Phi_a}e^{-i\epsilon{\cal H}}\cdots e^{-i\epsilon\l^a_1\Phi_a}e^{-i\epsilon
{\cal H}}e^{-i\epsilon\l^a_0\Phi_a}|p',q'\>\nonumber\\
&&\hskip2cm \times\,f(\epsilon\l_{N})\cdots f(\epsilon\l_0)\,\delta\epsilon\l_{N}\cdots\delta\epsilon\l_0\;.  
\en
Now we insert the coherent-state resolution of unity at appropriate places to find that the previous 
expression may also be given by
 \bn &&\hskip-1.5cm\lim\int\<p_{N+1},q_{N+1}|
e^{-i\epsilon\l^a_{N}\Phi_a}|p_N,q_N\>\prod_{l=0}^{N-1}\<p_{l+1},q_{l+1}|e^{-i\epsilon
{\cal H}}e^{-i\epsilon\l^a_{l}\Phi_a}|p_l,q_l\>\nonumber\\   &&\times 
[\prod_{l=1}^{N}d\mu(p_l,q_l)\,f(\epsilon\l_l)\,\delta\epsilon\l_l]\,f(\epsilon\l_0)\,\delta\epsilon\l_0\;.  \en
Following the pattern set in Sec.~2, this last expression may be readily turned into a formal coherent-state
 path integral given by
\bn \int\exp\{i\tint[p_j{\dot q}^j-H(p,q)-\l^a\phi_a(p,q)]\,dt\}\,\D\mu(p,q)\D E(\l)\;,  \en
where $E(\l)$ is a (possibly complex) measure designed so as to insert the projection operator $\E\,$ at 
every time slice. The integral over $E(\l)$ may be unity, but this normalization is not mandated as in the 
case of first-class constraints that form a closed Lie algebra. Also, unlike the case of the first-class 
constraints, we observe that the measure on the Lagrange multipliers is fixed. This usage of the Lagrange 
multipliers to ensure that the quantum system remains within the quantum constraint subspace is rather 
like their usage in the classical theory to ensure that the system remains on the classical constraint 
hypersurface. Thus it is not surprising that a fixed integration measure emerges for the Lagrange 
multipliers. On the other hand, it is also possible to use the measure $E(\l)$ in the case of {\it first}-class 
constraints as well; this would be just one of the acceptable choices for the measure $C(\l)$ designed to 
put at least one projection operator $\E\,$ into the propagator.

In summary, we have established the equality of the three expressions
\bn &&\hskip-1cm\<p'',q''|\E\;e^{-iT(\E\;{\cal H}\E\;)}\E\,|p,q\>\nonumber\\
&&=M''M'\int\exp\{i\tint[i\<p,q|\E\,\frac{d}{dt}|p,q\>/\<p,q|\E\,|p,q\>\nonumber\\
    &&\hskip2.3cm-\<p,q|\E\,{\cal H}\E\,|p,q\>/\<p,q|\E\,|p,q\>]\,dt\}\,\D_E\mu(p,q)\nonumber\\
&&= \int\exp\{i\tint[p_j{\dot q}^j-H(p,q)-\l^a\phi_a(p,q)]\,dt\}\,\D\mu(p,q)\D E(\l)\;.  \en
This concludes our derivation of path integral formulas for general constraints. Observe that we have not 
introduced any $\delta$-functionals, nor, in the last expression, reduced the number of integrations or the 
domain of integration in any way. 

As examples in the next section will show this formulation is suitable to discuss second-class constraints 
in a quite natural fashion.
\section{Examples of Second-Class Constraints}
\subsubsection*{First example of second-class constraints}
Consider the example with a single degree of freedom determined by the classical action
 \bn I=\tint[p{\dot q}-\half p^2-\quarter q^4-\l(p-1)-\xi(q-2)]\,dt\;,\en
where $\l$ and $\xi$ denote Lagrange multipliers. The stationary equations of motion read
 \bn  && {\dot q}=p+\l\;,\hskip1.5cm{\dot p}=-q^3-\xi\;,\nonumber\\
&&\hskip.5cm p=1\;,\hskip2cm q=2\;.  \en
It follows that $\l=-p=-1$ and $\xi=-q^3=-8$ in order that the constraints are satisfied for all time. Since 
$\{q-2,p-1\}=1$, we are dealing with second-class constraints because their Poisson bracket does not 
vanish on the constraint hypersurface.

Our path integral formulation is based on (101) and (106). In order to determine the projection operator 
$\E\,$ in this case we seek normalized states $|\phi\>$ for which 
\bn  \<\phi|(P-1)^2+(Q-2)^2|\phi\>={\rm minimum}\;\;\;(\propto\hbar)\;. \en
As is well known there is only one state that satisfies this requirement, and that state is the coherent state 
$|p=1,q=2\>=|1,2\>$ as defined by (19). Consequently, we choose $\E\,=|1,2\>\<1,2|$, namely, the 
one-dimensional projection operator onto this minimum uncertainty state. The theory of Weyl operators 
leads to the representation for $\E\,$ given by
\bn  |1,2\>\<1,2|=\int e^{-i\l(P-1)-i\xi(Q-2)}\,e^{-(\l^2+\xi^2)/4}\,d\l\,d\xi/(2\pi)\;. \en
Observe in this case that $\tint\exp[-(\l^2+\xi^2)/4]\,d\l\,d\xi/(2\pi)=2$, rather than unity as would be the 
case for a normalized measure. 

Having fixed our choice for $\E\,$ we may proceed to study the consequences of (101) and (106). In 
particular, (101) immediately reduces to 
\bn  \<p'',q''|1,2\>\,\<1,2|p',q'\>\,e^{-iH(1,2)T}\;.  \en
We now show that the formal expression (106) yields the same result. Observe that
\bn &&\hskip-2cm\tint\{[i\<p,q|\E\,\frac{d}{dt}\E\,|p,q\>]/\<p,q|\E\,|p,q\>\}\,dt\nonumber\\
&&=-\tint\{ {\rm Im\,}[\<1,2|p,q\>]^{-1}\frac{d}{dt}\<1,2|p,q\>\}\,dt\nonumber\\
&&={\rm Im\,}[\ln(\<p'',q''|1,2\>)+\ln(\<1,2|p',q'\>)]\;. \en
Likewise
\bn \<p,q|\E\,{\cal H}\E\,|p,q\>/\<p,q|\E\,|p,q\>=\<1,2|{\cal H}|1,2\>=H(1,2)\;.  \en
Thus the evaluation of the action in (106) for any path reads
 \bn {\rm Im\,}[\ln(\<p'',q''|1,2\>)+\ln(\<1,2|p',q'\>)]-H(1,2)T\;,  \en
which is {\it independent} of the particular path since it only depends on the fixed end points and the 
interval of time. Therefore, the integration of the path integral measure receives no weighting from the 
action. However, the integration at each time slice involves
 \bn  \tint\<p,q|\E\,|p,q\>\,d\mu(p,q)=\tint|\<1,2|p,q\>|^2\,d\mu(p,q)=1\;.\en
Hence, the result of the path integration (106) is given by
 \bn &&\hskip-1cm|\<p'',q''|1,2\>|\,|\<1,2|p',q'\>|\nonumber\\
&&\hskip.8cm\times\exp\{i{\rm Im\,}[\ln(\<p'',q''|1,2\>)+\ln(\<1,2|p',q'\>)]-iH(1,2)T\}\nonumber\\
&&=\<p'',q''|1,2\>\<1,2|p',q'\>e^{-iH(1,2)T}  \en
just as we found for (101). Observe that it has not been necessary to eliminate this second-class constraint, 
but that it can be dealt with in the context of a properly defined path integral.
\subsubsection*{Second example of second-class constraints}
As our second example we return to the (hyper)spherical constraint $q^2=1$ discussed in Sec.~6. If one 
deals with dynamics compatible with the constraint, e.g., for a classical Hamiltonian given in a natural 
notation by $(p\wedge q)^2$, then the analysis falls into the first-class constraint category. However, we 
will deliberately adopt an {\it in}compatible dynamics, and instead consider the system defined by the 
action functional
 \bn  I=\tint[p\cdot{\dot q}-\half p^2-\l(q^2-1)-\xi\,p\cdot q]\,dt\;, \en
where we have again used a standard vector and scalar product notation, as well as introduced a second 
Lagrange multiplier and constraint to ensure that the system remains on the configuration constraint 
hypersurface despite the tendency of the Hamiltonian to take one away from that surface. Note that the 
Poisson bracket of the two constraints is $\{(q^2-1),p\cdot q\}=2q^2$, which does not vanish on the 
constraint hypersurface. 

With regard to the quantization of this system, we confine our attention to a discussion of the projection 
operator $\E\,$. Just as in the previous example we seek normalized states $|\phi\>$ that satisfy
\bn \<\phi|(Q^2-1)^2+\quarter(P\cdot Q+Q\cdot P)^2|\phi\>={\rm minimum}\;\;\;(\propto \hbar)\;. \en
Whenever $J\geq2$, there are many states that satisfy this minimum. In a Schr\"odinger representation 
any state of the form $\zeta(r)f({\rm angles})$, where $r$ denotes the radius, $\zeta$ denotes a specific, 
fixed state [minimizing (122)], and $f$ is a general function of the ``angles'' ($\equiv 
x_j/r\equiv\gamma_j$), will satisfy this relation (we do not pursue the specific form of $\zeta$ further). 
The reproducing kernel including the projection operator onto the constraint subspace is given by
\bn &&\hskip-1cm\<p'',q''|\E\,|p',q'\>=\tint\eta^*(r\gamma-q''))e^{-ir p''\cdot\gamma}\,\zeta(r)r^{J-
1}\,dr\nonumber\\
&&\hskip2.3cm\times\tint\zeta^*(s)e^{isp'\cdot\gamma}\eta(s\gamma-q')s^{J-1}\,ds\nonumber\\
&&\hskip2.3cm\times2\delta(1-\gamma^2)\,\Pi\,d\gamma_j\;.  \en
In this expression $r$ and $s$ are positive radial coordinates, and $\gamma_j$ is a unit vector in the 
direction of the vector $\bf r$ and $\bf s$; note that these vectors have a common direction but 
independent lengths. It is straightforward to confirm that this expression is a reproducing kernel and 
satisfies
\bn  \<p'',q''|\E\,|p',q'\>=\tint\<p'',q''|\E\,|p,q\>\<p,q|\E\,|p',q'\>\,d\mu(p,q)\en
when integrated over $\ir^{2J}$ provided that
\bn  \tint|\zeta(r)|^2r^{J-1}\,dr=1\;,  \en
which we adopt as our normalization criterion for $\zeta$.

It is interesting to observe how the projection operator $\E\,$ depends implicitly on the form of the kinetic 
energy ($p^2/2$) and the vector $\zeta$ would in general change if the momentum dependence of the 
kinetic energy were to change. In addition, the configuration space constraint hypersurface also enters into 
the definition of $\zeta$, hence $\E\,$, through the choice of the second constraint equation. In principle, 
the method sketched here for a (hyper)spherical configuration constraint hypersurface can be extended to 
a general configuration constraint hypersurface without any special symmetry.
\section{Conclusions}
The principal concept underlying this paper consists of a reassessment of the usual measure used for 
Lagrange multipliers in the phase-space path integral quantization of dynamical systems with constraints. 
Rather than choosing a measure leading to $\delta$-functionals of the classical constraints, we have found 
alternative measures that enforce the appropriate quantum constraints. Equations (65), (66), and (70) 
constitute our main results for closed first-class constraints, while Eq.~(110) summarizes our main results 
for general constraints. In deriving these equations we have developed well-defined lattice expressions 
asserting that, with the appropriate meaasure for the Lagrange multipliers, the formal phase-space path 
integrals lead to the indicated and well-defined operator expressions. We note that our procedures: (i) 
apply to first- and second-class constraints; (ii) apply to constraints with discrete or continuous spectrum; 
and (iii) lead to a reduced Hilbert space but generally employ the same number of integration variables 
and integration domain as for the unconstrained system. On the other hand, our procedures: (i) do not 
introduce $\delta$-functionals of the classical constraints; (ii) do not require auxiliary conditions (i.e., a 
gauge choice) for first-class constraints nor the elimination of variables for second-class constraints; and 
(iii) do not introduce potentially ambiguous determinants.

Although a parallel discussion to the one presented here may well be developed for conventional phase-
space path integrals, we have chosen to use coherent states and associated coherent-state path integrals 
because of their inherently superior compatibility with general canonical coordinate transformations. A 
discussion of how such coordinate transformations are implemented and incorporated will be presented 
elsewhere \cite{jrk}.

The question naturally arises whether the present formulation leads to results that are identical with or 
that are different than results obtained by more conventional treatments. This is especially of interest in 
the extension of our methods to an infinite number of degrees of freedom relevant for field theory. On the 
one hand, the use of different phase-space path-integral measures points in the direction of possibly 
different results. On the other hand, both the original and the alternative measures of this paper are 
designed, in principle, to achieve the same end, namely, enforcing the constraints. This fact suggests 
possible equivalence of the results. It appears that only a detailed study of specific examples will clarify 
when the results of the various approaches will agree and when they will disagree.

\section*{Acknowledgements}
Thanks are expressed to J. Govaerts for an illuminating conversation regarding constraints. In addition, 
thanks are expressed to S. Shabanov, W. Tom\'e, G. Tulsian, G. Watson, and B. Whiting for numerous 
discussions.

\end{document}